\documentclass[%
 reprint,
 amsmath,amssymb,
 aps,
]{revtex4-1}
\usepackage[usenames, dvipsnames]{color}
\usepackage[autostyle]{csquotes}
\usepackage{pgfplots}
\usepackage{tikz}
\usepackage{graphicx}
\usepackage{dcolumn}
\usepackage{bm}

\begin{document}

\preprint{APS/123-QED}

\title{Affine Inflation}

\author{Hemza Azri}
 \email{hemzaazri@iyte.edu.tr}
\author{Durmu\c{s} Demir}%
 \email{demir@physics.iztech.edu.tr}
\affiliation{%
 Department of Physics, \.{I}zmir Institute of Technology, TR35430
\ \.{I}zmir, Turkey
}%

\date{\today}

\begin{abstract}
Affine gravity, a gravity theory based on affine connection with no notion of metric, supports scalar field dynamics only if scalar fields have non-vanishing potential. The non-vanishing vacuum energy ensures that the cosmological constant is non-vanishing. It also ensures that the energy-momentum tensor of vacuum gives the dynamically generated metric tensor. We construct this affine setup and study primordial inflation in it.  We study inflationary dynamics in affine gravity and general relativity, comparatively. We show that non-minimally 
coupled inflaton dynamics can be transformed into a minimally-coupled one with a modified potential. We also show that there is one unique frame in affine gravity, as opposed to the Einstein and Jordan frames in general relativity. Future observations with higher accuracy may be able to test the affine gravity.
 
\begin{description}
\item[Keywords]
Inflation, affine gravity, affine connection, vacuum energy, non-minimal coupling,\\ conformal transformation, conformal frames.
\end{description}
\end{abstract}

\pacs{Valid PACS appear here}
\maketitle


\section{Introduction}
\label{sec:introduction}

Inflation, exponential expansion of the early universe to facilitate its flatness and
homogeneity properties, rests on negative-pressure sources like
vacuum energy or slow-moving scalar fields \cite{guth,linde1,albrecht,linde2}. 
This conceptional idea gives us also the origin of the nearly scale-invariant spectrum 
of the cosmological perturbations. These predictions are tested at some level
by the anisotropy of the Cosmic Microwave Background (CMB) radiation 
as well as the Large Scale Structure (LSS) galaxy surveys \cite{planck}.

In the modern view, the basic idea of inflation is to postulate the existence of 
a scalar field, named \enquote{inflaton}, which fills a region existed in the 
early stage of the universe. This field is supposed to start with values slightly
larger than Planck mass, and lead to inflated domains. The inflationary dynamics 
have been studied mainly in  the metrical gravity (the general relativity (GR)).

In GR, which is the purely metrical theory of gravity, 
scalar fields can be coupled minimally and  non-minimally 
to gravity. In the first, the inflaton is coupled directly to the 
metric tensor and the inflationary regime is attained for 
the standard slow roll conditions applied to the scalar field. 
In this framework, inflationary models differ from each other
in the potential of the scalar field \cite{guth,linde1,albrecht,linde2}. 
Observations of density perturbations have severly constrained
these models. In view of this, generalizations to non-minimal 
coupling  have been proposed in the literature \cite{fakir,fakir2,kaiser,komatsu,futamase,makino},
 including the standard model Higgs boson as an inflaton \cite{bezrukov}. 
The non-minimal coupling $\xi$ enters into the theory as $\xi \phi^{2}\mathcal{R}$,
where $\phi$ is the inflaton and ${\mathcal{R}}$ the scalar curvature.   
 
The minimal and  non-minimal couplings both are studied in the GR, where
metric tensor is the fundamental variable. This is precisely the structure 
we observe at large distances. However, the spacetime structure 
may be different to start with in the early universe. In other words,
the metrical description of the GR might have arisen dynamically as 
the universe evolves. To this end, the affine gravity (AG) \cite{eddington,kijowski1,kijowski2,demir1,azri1,azri2,poplawski}, based 
solely on connection with no notion of metric, stands out a viable 
framework to study. The AG framework necessitates scalar fields to 
have nonvanishing potentials, and thus, studying inflation in the
AG is important by itself. We find that the non-zero vacuum energy
dynamically leads to metric tensor as its energy-momentum tensor.
This metric tensor is the consequence of the structure of the affine actions 
where the kinetic and the potential energies of scalars come out not
in addition but in division. We will study salient consequences
of this novel structure, and apply our findings to inflationary 
epoch as a concrete testbed. We will show how a non-minimally 
coupled scalar can be turned into a minimally-coupled one 
in the AG by a field redefinition.  We will study cosmological
inflationary parameters in affine inflation (AfI) as functions
of the  non-minimal coupling parameter $\xi$, and compare them 
with the predictions of the GR. 

The paper is organized as follows. In section~\ref{sec:level2},
we discuss minimally-coupled scalar field in the GR and the AG,
and reveal the differences and similarities between the two. We
show therein how metric tensor arises dynamically in the AG
and how it relates to the energy-momentum tensor of the 
vacuum. In section~\ref{sec:level3}, we extend our analysis 
to non-minimally coupled scalar fields and study again 
the GR and AG comparatively. Therein we point out 
an interesting property in that in the AG a non-minimally
coupled scalar field can be transformed into a 
minimally-coupled one by a field redefinition. (This 
is achieved in the GR by a conformal transformation of
the metric plus field redefinition.)  In section~\ref{sec:level4},
we apply our findings on scalar field dynamics to primordial
inflation. We study in detail basic inflationary parameters
in the AG and the GR, and depict our results in 
tables and plots. In  section~\ref{sec:level5} we conclude.

\section{\label{sec:level2} Minimally Coupled Scalar Field}

\subsection{\label{sub2} GR perspective}
The spacetime of GR is equipped with a metric tensor $g_{\mu\nu}$ which makes the notions
of distances and angles possible, and also forms the invariant volume via  $\sqrt{-g}$ factor.
In this theory, gravity-scalar field coupling is described by the following action
\begin{eqnarray}
S^{(1)}_{\text{GR}}=\int d^{4}x \sqrt{-g}\left[\frac{M_{Pl}^{2}}{2} \mathcal{R}\left(g \right)-\frac{1}{2} g^{\mu\nu}\partial_{\mu}\phi\partial_{\nu}\phi-V\left(\phi\right) \right] \nonumber \\
\label{gr minimal action}
\end{eqnarray}
where $\mathcal{R}\left(g\right)$ is the Ricci scalar curvature and $V\left(\phi\right)$ is the potential 
associated with the scalar field $\phi$. The reduced Planck mass relates to Newton's constant $G_N$ as 
$M^{2}_{Pl}=\left(8\pi G_{N} \right)^{-1}$.

The theory (\ref{gr minimal action}) including the celebrated Einstein-Hilbert action 
is based on the metric tensor $g_{\mu\nu}$ as a fundamental quantity. The GR then is a \textit{purely metric} 
theory of gravity. The gravitational equations  are then given by
\begin{eqnarray}
\label{einstein equations1}
M_{Pl}^{2}G_{\mu\nu}\left(g\right)=
\partial_{\mu}\phi \partial_{\nu}\phi-\frac{1}{2}g_{\mu\nu}\left(\partial\phi\right)^{2}
-g_{\mu\nu}V\left(\phi\right),
\end{eqnarray}
where $G_{\mu\nu}\left(g\right)$ is the Einstein's tensor constructed 
from $g_{\mu\nu}$ and the right-and side is the energy momentum tensor of 
the scalar field
\begin{eqnarray}
\label{energy momentum1}
T^{\phi}_{\mu\nu}= \partial_{\mu}\phi \partial_{\nu}\phi-\frac{1}{2}g_{\mu\nu}\left(\partial\phi\right)^{2}
-g_{\mu\nu}V\left(\phi\right).
\end{eqnarray}
The dynamics of the scalar field $\phi$ is described by 
the following equation derived from (\ref{gr minimal action}) by varying with respect to $\phi$
\begin{eqnarray}
\Box \phi -V^{\prime}\left(\phi\right)=0.
\end{eqnarray}
where prime stands for differentiation with respect to $\phi$. 

The scalar-tensor action in (\ref{gr minimal action}) sets the minimally-coupled 
scalar field dynamics. It possesses two important properties:
\begin{enumerate}
\item As in the case of the flat spacetime action, 
kinetic terms (derivatives) of the scalar field and 
potentials appear in the action in the same line as a sum of two terms.
\item As a result of the first property, all potentials $V\left(\phi\right)$ (zero or non-zero)
are admissible, and in the vacuum where $\phi=\phi_{min}$ one can optionally
set $V\left(\phi\right)=0$ or leave it non-zero depending on the model. 
\end{enumerate}
Next, we will consider the purely affine theory where the metric tensor 
is absent and see that these two properties no longer hold.

\subsection{\label{sub2} AG perspective}
This geometry possesses only affine connection. This
is all we need to define curvature. There is 
no metric tensor to start with; gravity is 
purely affine. 

A real scalar field $\phi$ with scalar potential 
$V(\phi)$ in affine spacetime of Ricci curvature $\mathcal{R}_{\mu\nu}\left(\Gamma\right)$ 
is governed by the action
\begin{eqnarray}
\label{action1}
S_{\text{AG}}^{(1)}=
\int d^{4}x\frac{\sqrt{{\mbox{Det}} \left[ M_{Pl}^{2}\mathcal{R}_{\mu\nu}\left(\Gamma\right)
-\partial_{\mu}\phi \partial_{\nu}\phi\right]}}{V\left(\phi\right)}, \\
\nonumber
\end{eqnarray}
wherein the connection $\Gamma^{\lambda}_{\mu\nu}$ is taken symmetric.  

The AG model in (\ref{action1}) is the simplest form of a pure affine theory of gravity coupled to a scalar field. 
This theory was shown to be equivalent to general relativity for
$V\left(\phi\right)=m^{2}\phi^{2}/2$, where the metric tensor arises as the momentum canonically 
conjugate to the connection \cite{kijowski1}. This proof can be straightforwardly 
extended to a general potential $V\left(\phi\right)$ \cite{kijowski2,azri2}.

Unlike the action (\ref{gr minimal action}) of GR, the AG action assumes the two properties below:
\begin{enumerate}
\item The derivatives of the scalar field (kinetic part) enters the dynamics along with the 
curvature tensor. They both appear in the determinant needed for invariant volume element. 
The non-derivative parts of the scalar field (potential 
part) appear in the denominator not to add to but to divide the kinetic part. 
\item The action (\ref{action1}) is then singular at $V(\phi) = 0$. This means that 
the scalar field must always have a non-zero potential energy. If $\phi = \phi_{\min}$ 
is the value of the scalar field for which $V(\phi)$ attains its minimum and if $V(\phi_{min}) \neq 0$ then 
theory makes sense, physically. In general, $\phi_{min}$ is constant (it may be zero) and hence $V(\phi_{min})$ 
is the vacuum energy.
\end{enumerate}

In the following, we will generate metric and its dynamical equations (Einstein field equations) through the 
action (\ref{action1}) by utilizing its above-mentioned properties. The important point here is that the potential energy, which must have a non-vanishing part always, is nothing but the energy-momentum tensor of vacuum, and it creates by itself a notion of metric. (In fact, even in the GR, metric can well be interpreted as the energy-momentum tensor of vacuum \cite{demir2}.) In this sense, affine spacetime filled with vacuum energy $V(\phi_{min})$ provides a very simple background which turns out to be the maximally symmetric spacetime (see the discussion at the end of this section and at the beginning of Sec~\ref{sec:level4}.)

This non-vanishing vacuum energy, speaking covariantly, implies the existence of a vacuum energy-momentum tensor, $T_{\mu\nu}$. It is a non-vanishing, invertible rank two tensor giving a covariant description of the vacuum energy. It is implicitly contained in the affine spacetime, and acts as a \enquote{dimensionful} metric tensor by the nature of vacuum. With non-singular inverse $(T^{-1})^{\lambda \rho}$, it defines the Levi-Civita connection
\begin{eqnarray}
\label{GammaT}
{}^{T}\Gamma^{\lambda}_{\,\,\,\mu \nu} = \frac{1}{2} (T^{-1})^{\lambda \rho} \left(\partial_{\mu} T_{\nu \rho} + \partial_{\nu} T_{\rho \mu} - \partial_{\rho} T_{\mu\nu}\right)
\end{eqnarray}
with respect to which 
\begin{eqnarray}
\label{compatibility}
\nabla^{T}_{\mu} T_{\alpha\beta} = 0. 
\end{eqnarray}
naturally holds.

To reveal more the structure of this energy-momentum tensor, one can also note that 
the identity tensor $\delta^{\mu}_{\nu}$ is inherently contained in affine spacetime, and thus, $T_{\mu\nu}$ can be incorporated in its mixed form as
\begin{eqnarray}
\label{vacuum tensor}
T^{\mu}_{\nu} && \equiv V(\phi_{min}) \delta^{\mu}_{\nu} \\
&&= V(\phi_{min})  T_{\nu\alpha} (T^{-1})^{\alpha\mu}. \nonumber
\end{eqnarray}
This manifests itself as part of the affine spacetime. It does not arise from raising the indices of the tensor $T_{\mu\nu}$ though it will do so when the metric tensor is defined through  $T_{\mu\nu}$ (see the equation (\ref{metric}) below.) 

Accordingly, $T_{\mu\nu}$ is essentially a \enquote{dimensionful} metric tensor. In a sense, the vacuum sets a metrical geometry. However, one thing remains
in that it is necessary to generate $T_{\mu\nu}$ dynamically from the 
equations of motion. (See the discussions in \cite{demir2}.)

The equation of motion arising from the affine action (\ref{action1}) takes the form
\begin{widetext}
\begin{eqnarray}
\nabla_{\mu} \left\lbrace \frac{\sqrt{{\mbox{Det}} \left[ M_{Pl}^{2}\mathcal{R}_{\mu\nu}-\partial_{\mu}\phi \partial_{\nu}\phi\right]}}{V\left(\phi\right)} 
\left(\left(M_{Pl}^{2}\mathcal{R}-\partial\phi \partial\phi\right)^{-1}\right)^{\alpha \beta}\right\rbrace = 0, \label{dynamical equation1}
\end{eqnarray}
\end{widetext}
where the covariant derivative operator $\nabla_{\mu}$ is with respect to the arbitrary affine connection $\Gamma$ that defines the Ricci tensor. 
Now, by taking into account the property (\ref{compatibility}), the last equation of motion is solved as
\begin{eqnarray}
\label{gravitational equations1}
M_{Pl}^{2}\mathcal{R}_{\mu\nu}-\partial_{\mu}\phi \partial_{\nu}\phi
=\left(\frac{V(\phi)}{V(\phi_{min})}\right)T_{\mu\nu},
\end{eqnarray}
so the affine connection coincides with ${}^{T}\Gamma^{\lambda}_{\mu\nu}$ given in (\ref{GammaT}).
This is in a sense the {vacuum connection}, the connection that is generated by 
the vacuum stress energy tensor.
Here, the metric tensor of GR is nothing but a tensor $g_{\mu\nu}$ where its existence is guaranteed by the non-zero vacuum energy via
\begin{eqnarray}
\label{metric}
g_{\mu\nu}=T_{\mu\nu}/V\left(\phi_{min}\right).
\end{eqnarray}
Clearly, this tensor is defined only for $V\left(\phi_{min}\right) \neq 0$, the condition that makes the theory derived 
from the action (\ref{action1}) factual. To that end, indices raising, lowering and contraction 
of tensors take their standard form by this metric tensor. As a result, the gravitational equations 
(\ref{gravitational equations1}) can be easily brought into the form of 
Einstein's equations (\ref{einstein equations1}).

Unlike the metric tensor (and its Levi-Civita connection) 
which is usually supposed to be resulted from the dynamical equation (\ref{dynamical equation1}) as a constant of integration, 
the vacuum stress energy tensor (and hence the vacuum connection) is given a priori in the affine spacetime which translates a non-zero minimum potential energy of 
matter into a metrical geometry \cite{demir2}.

It must be emphasized here that the structure of the vacuum given by the stress-energy connection (\ref{GammaT}), and the energy-momentum tensor (\ref{vacuum tensor}) is not restricted to local minima of the potential. All one needs is a non-zero primordial piece in $V(\phi)$, which can be defined as minimum value of $V(\phi)$ corresponding to $\phi_{min}$. This constant value saves the action (\ref{action1}) from going singular. In general, the potential $V(\phi)$ is model dependent and its minimum can be reached even asymptotically. This does not affect the definition of the Levi-Civita connection (\ref{GammaT}).

With all these at hand, variation of the action with respect to the scalar field $\phi$ 
leads to the dynamical equation of motion of $\phi$
\begin{eqnarray}
\label{scalar field equation1}
\Box \phi -V^{\prime}\left(\phi \right)=0,
\end{eqnarray}
where we have used the solution (\ref{gravitational equations1}) to get the operator $\Box$. 

As we have seen, the equations of motion (\ref{gravitational equations1}) and 
(\ref{scalar field equation1}) derived from the affine action (\ref{action1}) are 
already familiar from the field equations of GR derived from Einstein-Hilbert action, 
where the scalar field is coupled minimally. This shows that coupling matter in 
AG through action (\ref{action1}) is {equivalent} to minimal coupling in GR. 
A summary of this comparison is given in table~\ref{tab:table1}.

\begin{table}[h]
\begin{ruledtabular}
\begin{tabular}{cccc}
\textrm{}&
\textrm{GR}&
\multicolumn{1}{c}{\textrm{AG}}&
\textrm{AG vs GR}\\
\colrule
\, & \, & \, & \, \\
Fundamental \\ quantity & Metric & Connection & \,\\
\, & \, & \, & \, \\
Action & (\ref{gr minimal action}) & (\ref{action1}) & Equivalent\\
\, & \, & \, & \, \\
Field \\ equations & (\ref{einstein equations1}) & (\ref{einstein equations1}) & \,\\
\end{tabular}
\end{ruledtabular}
\caption{\label{tab:table1}%
AG vs GR for Minimal coupling case.
}
\end{table}

Generation of the metric 
tensor can be understood through the fact that equation (\ref{dynamical equation1}), with $\phi=\phi_{min}$, 
has a solution of the form $\mathcal{R}_{\mu\nu}=\Lambda g_{\mu\nu}$, which defines metric tensor as 
in equation (\ref{metric}). For $\Lambda=0$, curvature vanishes, metric becomes irrelevant and metric description fails. 
Thus the Eddington solution with $\Lambda \propto V\left(\phi_{min}\right)$ defines the metric tensor \cite{eddington}.

The formalism presented here goes beyond the original Eddington approach, where matter fields are not included. It produces dynamically the metric and the cosmological constant as constants of integration. In our case, however, the affine spacetime is filled with a scalar matter to start with. In this sense, the theory described by the action (\ref{action1}) improves on Eddington's approach. The vacuum $V(\phi_{min})$ manifests itself as the non-zero energy required for elucidating the metric tensor. In other words, the metric tensor, though an integration constant as in Eddington's approach, can be structured in our case as the energy-momentum tensor of vacuum.

\section{\label{sec:level3} Non-Minimally Coupled Scalar Field}

\subsection{GR case}
Non-minimal coupling in GR corresponds to a direct coupling between the 
scalar field and the curvature scalar.  In this case, the action (\ref{gr minimal action}) is 
extended by adding an explicit interaction term between
$\phi$ and $\mathcal{R}\left(g \right)$ as follows
\begin{eqnarray}
\label{gr nonminimal action}
S_{\text{GR}}^{(2)}=S_{\text{GR}}^{(1)}+\int d^{4}x\sqrt{-g}\left(\frac{\xi}{2}\phi^{2}\mathcal{R}\left(g\right)\right),
\end{eqnarray}
where $\xi$ is a dimensionless parameter. It is then straightforward to obtain the gravitational field equations 
\begin{eqnarray}
\label{field equation nm gr}
M_{Pl}^{2}G_{\mu\nu}\left(g\right)=T_{\mu\nu}^{\phi}+\xi \nabla_{\mu}\nabla_{\nu}\phi^{2}-\xi \Box \phi^{2}g_{\mu\nu}-\xi \phi^{2}G_{\mu\nu}\left(g \right) \nonumber \\
\end{eqnarray}
where $T_{\mu\nu}^{\phi}$ is the energy momentum of the scalar field given in (\ref{energy momentum1}). Similarly, the 
equation of motion for the scalar field takes the form 
\begin{eqnarray}
\Box \phi -V^{\prime}\left(\phi \right)+\xi\phi \mathcal{R}\left(g\right)=0.
\end{eqnarray}
In consequence,  following properties concerning the form of the action and the equations of motion are worth noting:
\begin{enumerate}
\item As we see from the total action (\ref{gr nonminimal action}), the non-minimal coupling term 
$\xi \phi^{2} \mathcal{R}\left(g\right)$ appears in the theory as an additional term. 
This is a property of coupling to gravity in the pure metrical picture.

\item
Correction to the energy momentum tensor of scalar field due to non-minimal coupling has the following form
\begin{eqnarray}
\label{improved GR}
T_{\mu\nu}^{\text{GR}}= \xi \nabla_{\mu}\nabla_{\nu}\phi^{2}-\xi \Box \phi^{2}g_{\mu\nu}-\xi \phi^{2}G_{\mu\nu}\left(g \right).
\end{eqnarray}
The first two terms of this tensor arise here due to the {nonlinearity} of the action, they contain second derivatives of the metric tensor. This creates derivatives of the scalar field, 
and then the improved energy momentum tensor emerges as kinetic terms of matter. 
For a constant field $\phi=\phi_{0}$, these terms disappear leaving behind no contribution to 
the cosmological constant.
\end{enumerate}

Next we will study the corresponding non-minimal coupling in AG and see the differences.

\subsection{AG case}
Equivalence between the gravitational field equations that are derived from AG and GR actions 
in the minimal coupling case leads to the following questions:
\begin{itemize}
\item Is the gravity-scalar field coupling given in the action (\ref{action1}) minimal?
\item If yes, what is the generalization of this action to a non minimal case? 
Are the field equations derived from this new theory equivalent to the associated 
non-minimal case of GR?
\end{itemize}

Firstly, as we have seen so far, in GR, the invariant volume element, which is 
required for integration on spacetime, is independent of matter fields. It is
determined by  the scalar density $\sqrt{-g}$ of the metric tensor. Invariant quantities are 
then formed by matter fields contracted with this metric. 
However, in AG, the invariant integration measure explicitly involves 
the matter fields, as it is clear from the action (\ref{action1}). Thus, the comparison with the 
non-minimal case of GR may not be straightforward.
Here we propose a possible and simple generalization of action (\ref{action1}) as follows
\begin{eqnarray}
\label{action2}
S_{\text{AG}}^{(2)} = \int d^{4}x \frac{\sqrt{{\mbox{Det}}\left[\left(M_{Pl}^2 + \xi \phi^2\right)\mathcal{R}_{\mu\nu}\left(\Gamma\right) - \partial_{\mu}\phi\partial_{\nu}\phi \right]}}{V(\phi)}.\nonumber  \\
\end{eqnarray}
Construction of this affine action is performed by using kinetic terms of the matter fields and their
coupling terms to  the Ricci tensor. This automatically coincides with the action (\ref{action1}) for $\xi =0$.

Our aim in this paper is to study the gravitational dynamics and the dynamics of the scalar field 
which is {non-minimally} coupled to gravity in affine spacetime through action (\ref{action2}). 
To that purpose, it is important to shed light again on some points concerning the structure of this action
\begin{enumerate}
\item Unlike GR where the non-minimal coupling term in action (\ref{gr nonminimal action}) 
arises as an additional term,  the $\xi \phi^{2}\mathcal{R}_{\mu\nu}(\Gamma)$ interaction
in (\ref{action2}) is part of the invariant integration measure and does not come in an
additive action piece. 
\item
The theory becomes singular if at some values of $\phi$, the potential vanishes. This means
that there must be a primordial piece in $V(\phi)$. There is, however, an alternative view. It may be said that a constant $\phi$ defines complete absence of the scalar field (see equation (\ref{potential}) in Sec \ref{sec:level4}). The interesting point is
that the requisite primordial piece $V_0$ in the potential can 
be interpreted as $V(\phi_{min})$. 
\item
Needless to say, kinetic terms of matter fields vanish for a constant potential 
$V\left(\phi_{min} \right) \neq 0$. It is this structure of affine spacetime 
that accommodates the vacuum energy as an essential ingredient needed to forbid the 
singular behaviour of the theory.
\end{enumerate}

Action (\ref{action2}) is the simplest possible generalization of (\ref{action1}). First of all, this choice is structured as the one that gives the minimal form in (\ref{action1}) in the limit $\xi=0$. The action (\ref{action2}) maintains the same fundamental structure, in which the kinetic term (not modified here) takes part in defining the volume element (square root of the determinant) and the potential divides the volume element. The step taken here is to couple explicitly the field $\phi$ with the Ricci tensor, which is the only geometric quantity in the action. It is this form that goes beyond minimal coupling as it provides direct coupling between $\phi$ and the connection $\Gamma$. This can, of course, be generalized to more general forms like $\mathcal{F}(\phi)$ rather than $\xi \phi^{2}$. However, higher powers of $\phi$ are expected to be suppressed by the Planck mass.

Now, the dynamical field equations derived from action (\ref{action2}) take the form
\begin{widetext}
\begin{eqnarray}
\nabla_{\mu} \left\lbrace \sqrt{{\mbox{Det}}\left[\left(M_{Pl}^2 + \xi \phi^2\right)\mathcal{R} - \partial\phi\partial\phi\right]} \frac{\left(M_{Pl}^2 + \xi \phi^2\right)}{V(\phi)}
\left(\left(\left(M_{Pl}^2 + \xi \phi^2\right)\mathcal{R} - \partial\phi\partial\phi\right)^{-1}\right)^{\alpha \beta} \right\rbrace = 0, \label{dynamical equation2}
\end{eqnarray}
\end{widetext}
which can be integrated as 
\begin{eqnarray}
\label{gravitational equations2}
&&(M_{Pl}^2 + \xi \phi^2) \mathcal{R}_{\mu\nu} - \partial_{\mu}\phi \partial_{\nu} \phi\nonumber\\ &=& \left(\frac{V(\phi)}{V(\phi_{min})}\right) \left(\frac{M_{Pl}^2}{M_{Pl}^2 + \xi \phi^2}\right) T_{\mu\nu}.
\end{eqnarray}
Again, in terms of the metric tensor (\ref{metric}), equation (\ref{gravitational equations2}) is written as
\begin{eqnarray}
M_{Pl}^2 G_{\mu\nu}\left(g \right)
= \partial_{\mu}\phi \partial_{\nu} \phi -\frac{1}{2}g_{\mu\nu} \left( \partial \phi\right)^{2}
\nonumber \\
-g_{\mu\nu}\frac{V\left( \phi\right)}{\mathcal{F}\left(\phi\right)} -\xi \phi^{2}G_{\mu\nu}\left(g\right),
\label{einstein equations2}
\end{eqnarray}
where we have defined for brevity the function $\mathcal{F}\left(\phi\right)$ as follows
\begin{eqnarray}
\label{f(phi)}
\mathcal{F}\left(\phi\right)=1+\frac{\xi \phi^{2}}{M^{2}_{Pl}}.
\end{eqnarray}
These are the gravitational field equations resulting from the non-minimal 
coupling of the scalar field to gravity in affine spacetime. 
The right-hand side term of equation (\ref{einstein equations2}) 
is the generalized energy momentum tensor of the scalar field which can be written as
\begin{eqnarray}
\label{total energy momentum}
T_{\mu\nu}\left(\phi\right)=T_{\mu\nu}^{\phi}+T_{\mu\nu}^{\text{AG}}\left(\phi\right),
\end{eqnarray}
where $T_{\mu\nu}^{\phi}$ is the standard energy momentum tensor (\ref{energy momentum1}) 
derived from the minimal coupling case. The term $T_{\mu\nu}^{\text{AG}}$ is an improved energy momentum tensor
\begin{eqnarray}
\label{improved AG}
T_{\mu\nu}^{\text{AG}}=
\frac{\xi\phi^{2} }
{M^{2}_{Pl}+\xi \phi^{2}}
V\left(\phi\right)g_{\mu\nu}-\xi \phi^{2}G_{\mu\nu}\left(g\right).
\end{eqnarray}
Obviously, this quantity vanishes for $\xi=0$ -- the minimal coupling case. 

Now variation of the action (\ref{action2}) with respect to the 
scalar field $\phi$ leads to the following equation of motion
\begin{eqnarray}
\label{scalar field equation2}
\Box \phi -V^{\prime}\left(\phi \right)+\xi \phi \mathcal{R}\left(g\right)+\Psi\left(\phi\right)=0,
\end{eqnarray}
where we have used the identity (\ref{dynamical equation2}) and then (\ref{metric}).
Here the function $\Psi\left(\phi\right)$ is given by
\begin{align}
\Psi\left(\phi\right)&=
\frac{\xi\phi^{2}}{M^{2}_{Pl}+\xi\phi^{2}}V^{\prime}\left(\phi\right) \nonumber \\& 
-\left(\frac{2\xi\phi}{M^{2}_{Pl}+\xi\phi^{2}}\right) g^{\mu\nu}\nabla_{\mu}\phi\nabla_{\nu}\phi.
\end{align}
Equation (\ref{scalar field equation2}) implies the covariant conservation of the 
energy momentum tensor (\ref{total energy momentum});
\begin{eqnarray}
\nabla^{\mu}T_{\mu\nu}\left(\phi\right)=0.
\end{eqnarray}
This is a consequence of the general covariance of the affine action (\ref{action2}).

The last two terms of equation (\ref{scalar field equation2}) measure the deviation from the 
dynamics of the scalar field in the minimal coupling case (\ref{scalar field equation1}).
The AG dynamics has the following properties:
\begin{itemize}
\item The dynamics of the scalar field is equivalent to the prescription of GR for $\xi =0$. However, in the general case, the affine dynamics shows 
no equivalence to GR due to the presence of $\Psi\left(\phi\right)$. Like the improved energy momentum tensor of the scalar field (\ref{improved AG}) 
in the gravitational sector, the quantity $\Psi\left(\phi\right)$ might impose constraints on the propagation of matter fields in the curved background.
\item
The first term of the improved tensor (\ref{improved AG}) shows no dependence on the field derivatives, 
this is a consequence of the {linearity} of the affine action (\ref{action2}) where the fundamental quantity 
is an affine connection. Unlike the GR case, this term emerges in the theory as a potential term rather 
than derivatives of the field. For a general constant field $\phi=\phi_{0}$, the improved term does not 
vanish but rather creates a cosmological constant. Thus,
\begin{enumerate}
\item The first term of the improved energy momentum tensor (\ref{improved AG}) is the measure of 
shifts between AG and GR in the non-minimal coupling case and new observable effects if any would arise through it.
\item The same term is essential in AG and it may enable us to shed light on some new features of 
the cosmological constant both classically and quantum mechanically \cite{demir1,demir2,demir3,sakharov,naturalness}.
\end{enumerate}
\end{itemize}
AG vs GR is summarized in table~\ref{tab:table2} for the non-minimally coupled scalar fields.

We conclude this discussion by shedding light on an important point concerning the transformation between minimal and non-minimal coupling in GR and AG:
\begin{itemize}
\item
In GR, the transition between the two actions (\ref{gr minimal action}) and (\ref{gr nonminimal action}) is made using the familiar conformal 
transformations where both actions are considered as the same theory written in two different frames. 
The  Jordan and Einstein frames are described by \textit{two metric tensors} $g_{\mu\nu}$ and $\tilde{g}_{\mu\nu}$ which are related by
\begin{eqnarray}
\tilde{g}_{\mu\nu}=\mathcal{F}\left(\phi\right)g_{\mu\nu}.
\end{eqnarray}
The question then of which frame or which metric should be considered physical causes a serious ambiguity in GR.
\item However, in AG, no such frames make sense in the theory. In fact, there is a unique metric 
tensor given by (\ref{metric}) which has originated from the non-zero vacuum energy. 
In this setup, the transition from the non-minimal affine action (\ref{action2}) to the 
minimal affine action (\ref{action1}) is obtained only through field redefinition
\begin{eqnarray}
\label{field transformation}
d\varphi=\frac{d\phi}{\sqrt{\mathcal{F}\left(\phi\right)}}.
\end{eqnarray}
In terms of this new field, action (\ref{action2}) becomes
\begin{eqnarray}
\label{transformed action}
S_{\text{AG}}=
\int d^{4}x\frac{\sqrt{{\mbox{Det}} \left[ M_{Pl}^{2}\mathcal{R}_{\mu\nu}\left(\Gamma\right)
-\partial_{\mu}\varphi \partial_{\nu}\varphi\right]}}{U\left(\varphi\right)},
\end{eqnarray}
which describes a minimal coupled scalar field $\varphi$ in affine space with the potential
\begin{eqnarray}
U\left[\varphi\left(\phi\right)\right]=\frac{V\left(\phi\right)}{\mathcal{F}^{2}\left(\phi\right)}.
\end{eqnarray}
This interesting feature of AG is not restricted to single fields but it holds true also in multi-scalar theories \cite{azri-progress}. The main impact of the passage from the non-minimal to minimal coupling cases is the new interactions induced. The multi-scalar theories, for instance, can develop new interactions (even after diagonalizing their kinetic terms). It is therefore inferred that non-minimally coupled scalars can always be reduced to minimally-coupled scalars with modified interactions with other matter fields.

Action (\ref{transformed action}) will be the basis in our discussion of the inflationary regime in the following section.

\end{itemize}

\begin{table}[h]
\begin{ruledtabular}
\begin{tabular}{cccc}
\textrm{}&
\textrm{GR}&
\multicolumn{1}{c}{\textrm{AG}}&
\textrm{AG vs GR}\\
\colrule
\, & \, & \, & \, \\
Fundamental \\ quantity & Metric & Connection & \,\\
\, & \, & \, & \, \\
Action & (\ref{gr nonminimal action}) & (\ref{action2}) & Different\\
\, & \, & \, & \, \\
Field \\ equations & (\ref{field equation nm gr}) & (\ref{einstein equations2}) & \,\\
\end{tabular}
\end{ruledtabular}
\caption{\label{tab:table2}%
AG vs GR for Non-minimal coupling case.
}
\end{table}

\section{ \label{sec:level4} Affine Inflation}

The gravitational field equations (\ref{einstein equations2}) 
take a simpler form when $\phi=\phi_{min}$. 
This is the maximally symmetric vacuum case and it leads to Einstein's equations with a cosmological constant (CC)
\begin{eqnarray}
\label{einstein with cc}
M^{2}_{Pl}G_{\mu\nu}\left(g\right)=-\frac{V\left(\phi_{min}\right)}{\mathcal{F}^{2} \left(\phi_{min}\right)}g_{\mu\nu}.
\end{eqnarray}
Solution to this equation is the {maximally symmetric} de Sitter (anti-de Sitter) spacetime.
The non-zero CC is the requirement of the structure of the model proposed here and then the 
cosmological effects of this term is relevant to the purely affine theory. As we have shown in 
Sec \ref{sec:level2}, the necessity of non-zero CC is hidden in Eddington's approach and the 
equations (\ref{einstein with cc}) are equivalent to Eddington's equations \cite{eddington}.

The symmetry requirements of isotropy and homogeneity of space lead to 
Friedmann-Lema\^{i}tre models for the universe. These models naturally include de Sitter 
solution and those incorporating the cosmological constant like the one given in the present work \cite{luminet}.
The spacetime is described by one special case of these models; the flat Robertson-Walker metric
\begin{eqnarray}
\label{frw}
ds^{2}=-dt^{2}+a^{2}\left(t\right)d\vec{\textbf{x}} \boldsymbol{\cdot} d\vec{\textbf{x}},
\end{eqnarray}
where $a\left(t\right)$ is the scale factor.

The distribution of the scalar field in the universe may be described by its associated energy density and pressure respectively
\begin{eqnarray}
\label{density and pressure}
\rho\left(\phi \right)= \frac{1}{ \mathcal{F}\left(\phi \right)} \left(\frac{\dot{\phi}^{2}}{2} +\frac{V\left(\phi \right)}{\mathcal{F\left(\phi \right)}} \right) \\
p\left(\phi \right)= \frac{1}{\mathcal{F}\left(\phi \right)} \left(\frac{\dot{\phi}^{2}}{2} -\frac{V\left(\phi \right)}{\mathcal{F}\left(\phi \right)} \right).
\end{eqnarray}
Here we see that a quasi-de Sitter solution which requires $p\left(\phi \right)= -\rho \left( \phi \right)$ is possible for 
some slowly rolling fields. The CC case we discussed above is implicitly understood here for $\phi=\phi_{min}$.

Now the Hubble parameter $H$ satisfies the following equations that can straightforwardly be derived from 
the gravitational field equations (\ref{einstein equations2})
\begin{eqnarray}
\label{hubble1}
H^{2}=\frac{1}{3 M_{Pl}^{2} \mathcal{F}\left(\phi \right)} \left(\frac{\dot{\phi}^{2}}{2}
+\frac{V\left(\phi \right)}{\mathcal{F\left(\phi \right)}}  \right),
\end{eqnarray}
and
\begin{eqnarray}
\label{hubble2}
\dot{H}+H^{2}= -\frac{1}{3M_{Pl}^{2} \mathcal{F}\left(\phi \right)} \left(\frac{\dot{\phi}^{2}}{2} -\frac{V\left(\phi \right)}{\mathcal{F}\left(\phi \right)} \right).  \\
\nonumber
\end{eqnarray}
The existence of quasi-de Sitter solution where the Hubble parameter (\ref{hubble1}) is constant shows that an inflationary regime is possible in the theory.

Theories of inflation driven by scalar fields coupled non-minimally to gravity have been studied in great detail in pure metric 
gravity \cite{fakir,fakir2,kaiser,komatsu,futamase,makino}. The study is usually performed in both 
Jordan and Einstein frames where same predicted results are not guaranteed. Here we will apply the formalism 
developed so far in this paper to inflation and compare the results with those predicted by GR. 

Here, we adopt the following potential which satisfies the standard properties discussed in two previous sections
\begin{eqnarray}
\label{potential}
V\left(\phi\right)= V_0 + \frac{\lambda}{4}\left(\phi^{2}-v^{2}\right)^{2},
\end{eqnarray}
where the $v$ is the vacuum expectation value of $\phi$. 

The $V_{0}$ is non-zero; it saves the affine action (\ref{action2}) from going singular at $\phi=v$. In fact, assuming that all possible contributions to vacuum energy are incorporated into $V_{0}$ and the cosmological constant problem is somehow solved, we take $V_{0}\simeq m_{\nu}^4$ (since $V_{0}$ sets the cosmological constant as $\Lambda=V_{0}/M^{2}_{Pl}$). It is clear that non-vanishing 
of the vacuum energy ensures non-vanishing nature of the cosmological constant -- an observationally known fact. During the inflationary epoch, $V_{0}\sim m_{\nu}^{4}$ is too tiny to have any observable effect and it will be dropped in the analyses below. (Of course, in the vacuum $\phi=v$, $V_{0}$ is crucial.)

It is easier to study the inflationary regime using the physical field $\varphi$ given 
by (\ref{field transformation}). At that end, equation (\ref{field transformation}) 
is integrated straightforwardly to get
\begin{eqnarray}
\phi\left(\varphi\right)=\frac{M_{Pl}}{\sqrt{\xi}}\sinh\left(\frac{\sqrt{\xi}}{M_{Pl}}\varphi\right).
\end{eqnarray}
In spite of using the field $\varphi$ rather than $\phi$, 
the spacetime metric (\ref{frw}) remains unchanged and then, 
the physical field $\varphi$, satisfies the standard slow roll conditions
\begin{eqnarray}
\label{slow roll conditions}
\frac{\dot{\varphi}^{2}}{2} \ll U\left(\varphi\right),\,\, \frac{\ddot{\varphi}}{\dot{\varphi}} \ll
H,
\end{eqnarray}
for the following potential
\begin{eqnarray}
U\left(\varphi\right)\simeq \frac{\lambda}{4}\left[ \frac{M^{2}_{Pl}\xi^{-1}\sinh^{2}\left(\frac{\sqrt{\xi}}{M_{Pl}}\varphi\right)-v^{2}}
{1+\sinh^{2}\left(\frac{\sqrt{\xi}}{M_{Pl}}\varphi\right)}\right]^{2}.
\end{eqnarray}
Now, the Hubble parameter and the equation of motion of $\phi$ are written as
\begin{eqnarray}
H^{2}\simeq \frac{U\left(\varphi\right)}{3M^{2}_{Pl}},\,\, \text{and}\,\,3H\dot{\varphi}\simeq
-U^{\prime}\left(\varphi\right).
\end{eqnarray}
For large field $\varphi >M_{Pl}/\sqrt{\xi}$, the slow roll parameters take the following forms
\begin{eqnarray}
\label{slow roll parameters}
\epsilon=\frac{M^{2}_{Pl}}{2}\left(\frac{U^{\prime}}{U}  \right)^{2}
\simeq 128\xi \exp \left(-4\frac{\sqrt{\xi}}{M_{Pl}}\varphi \right) \\
\eta= M^{2}_{Pl} \left(\frac{U^{\prime\prime}}{U}  \right)
\simeq -32\xi \exp \left(-2\frac{\sqrt{\xi}}{M_{Pl}}\varphi \right) \\
\zeta^{2}= M^{4}_{Pl}\frac{U^{\prime\prime\prime}U^{\prime}}{U^{2}} \simeq
\left(32 \xi\right)^{2}\exp \left(-4\frac{\sqrt{\xi}}{M_{Pl}}\varphi \right).
\end{eqnarray}
These are equivalent to the results obtained from Palatini formalism \cite{bauer}.

The number of e-foldings is given by
\begin{align}
N&=\frac{1}{M^{2}_{Pl}}\int_{\varphi_{f}}^{\varphi_{i}}\frac{U\left(\varphi\right)}{U^{\prime}\left(\varphi\right)}d\varphi \nonumber \\ &
\simeq\frac{1}{32\xi}\left[\exp\left(2 \frac{\sqrt{\xi}}{M_{Pl}}\varphi_{i}\right)- 
\exp\left(2 \frac{\sqrt{\xi}}{M_{Pl}}\varphi_{f}\right)\right].& \label{efoldings}
\end{align}
Here the final field $\varphi_{f}$ corresponds to the end of inflation where the 
slow roll conditions (\ref{slow roll conditions}) break down, or $\epsilon \simeq 1$, 
and the initial field $\varphi_{i}$ is determined from the number of e-foldings $N$.

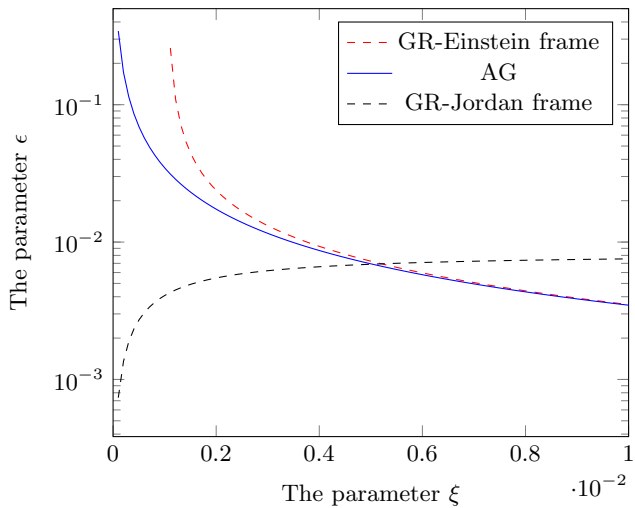
\begin{figure}[h]
\begin{tikzpicture}
\begin{axis}[
    axis lines = box,
    ymode=log,
    xmin=0,
    xmax=0.010,
    ymin=0,
    ymax=0.5,
    xlabel = The parameter $\xi$,
    ylabel = {The parameter $\epsilon$},
    legend pos=north east,
]
\addplot [
    domain=0:0.010,
    samples=100,
    color=red,
    style=dashed,
]
{32*x/((16*x*(60) - 1)*(16*x*(60) + 1))};
\addlegendentry{$\text{GR-Einstein frame}$}
\addplot [
    domain=0:0.010,
    samples=100,
    color=blue,
]
{1/(8*(60)^(2)*x)};
\addlegendentry{$\text{AG}$}
\addplot [
    domain=0:0.010,
    samples=100,
    color=black,
    style=dashed,
]
{8*x/(16*x*(60) + 1)};
\addlegendentry{$\text{GR-Jordan frame}$}
\end{axis}
\end{tikzpicture}
\caption{The slow roll parameter $\epsilon$ as a function of $\xi$. }
\label{fig:slow roll parameter}
\end{figure}

The slow roll parameters are evaluated at the value $\varphi$ when the scale of interest 
crossed the horizon during the inflationary phase, and they should remain smaller than one and 
then deviations of the spectrum of perturbations from scale invariant spectrum are small.
The smallness of the parameter $\epsilon$ is shown in Figure~\ref{fig:slow roll parameter} 
in terms of $\xi$. The parameter behaves as in GR only for large $\xi$.

Now, at first order the spectral index $n_{s}=1-6\epsilon+2\eta$ is written as
\begin{eqnarray}
n_{s}\simeq 1-\frac{3}{4\xi N^{2}}-\frac{2}{N}.
\end{eqnarray}
It has been shown that to first order, the non-minimal coupling 
in GR yields the following spectral index \cite{kaiser}
\begin{align}
\label{spectral index gr}
    n_{s}\simeq
\begin{cases}
    1-\frac{32\xi}{16\xi N-1},& \text{for } \phi_{f}^{2} \gg v^{2}\\
    1-\frac{16\xi\left(1+\delta^{2} \right)}{8\xi N\left(1+\delta^{2}\right)+\delta^{2}}         & \text{for } \phi_{f}^{2} \simeq v^{2}
\end{cases}
\end{align}
where $\delta^{2}=\xi v^{2}/M^{2}_{Pl}$.

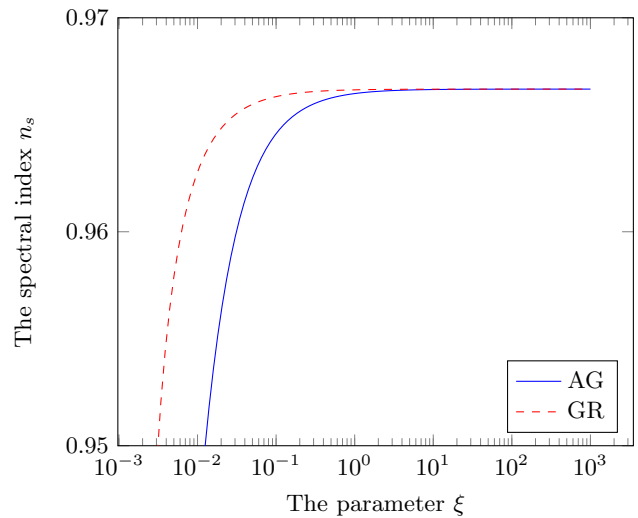
\begin{figure}[h]
\begin{tikzpicture}
\begin{axis}[
    xmode=log,
    axis lines = box,
    xlabel = The parameter $\xi$,
    ylabel = {The spectral index $n_{s}$},
    ymin=0.950,
    ymax=0.970,
    ytick={0.950,0.960,0.970},
    legend pos=south east,
]
\addplot [
    domain=0.003:10^(3),
    samples=100,
    color=blue,
]
{1-3/(4*x*(60)^(2))-2/(60)};
\addlegendentry{$\text{AG}$}
\addplot [
    domain=0.003:10^(3),
    samples=100,
    color=red,
    style=dashed,
    ]
    {1 - 32*x/(60*16*x - 1)};
\addlegendentry{$\text{GR}$}
\end{axis}
\end{tikzpicture}
\caption{First order spectral index $n_{s}$ in GR and AG for an e-foldings $N=60$. 
Planck result \cite{planck}, $0.960  \lesssim n_{s} \lesssim 0.970$, corresponds to 
$\xi \gtrsim 6.25 \times 10^{-3}$ for GR and
$\xi \gtrsim 3.12 \times 10^{-2}$ for AG.}
\label{fig: first order spectral index}
\end{figure}

The first order spectral index predicted by AG and GR for larger fields is depicted in Figure~\ref{fig: first order spectral index} 
in terms of the parameter $\xi$, for $N=60$. This comparison is made for $\phi \gg v$, 
where the potential behaves like $\phi^{4}$.

The observed value, $n_{s}\simeq 0.9655 \pm 0.0062$ \cite{planck}, is reached quickly, 
i.e, for smaller $\xi$ in GR than in AG. For large $\xi$, AG is closer to the observed values. 
A possible larger values of $\xi$ in AG may give rise to smaller ratios $\varphi/M_{Pl}$ 
even when $\sqrt{\xi}\varphi/M_{Pl}$ is large as it is required for the inflationary regime.

To second order, the spectral index $n_{s}$ depends explicitly on the third slow roll 
parameter $\zeta^{2}$ as follows \cite{liddle,stewart}
\begin{eqnarray}
\label{second order spectral index}
&n_{s}= 1-6\epsilon +2\eta +\frac{1}{3}\left(44-18c\right)\epsilon^{2}+\left(4c-14 \right)\epsilon \eta \nonumber \\
&+\frac{2}{3}\eta^{2}+\frac{1}{6}\left(13-3c \right)\zeta^{2},
\end{eqnarray}
where $c=4\left(\ln 2 +\gamma \right)\simeq 5.081$ and $\gamma$ being Euler's constant.

Detailed analysis in both Jordan and Einstein frames showed that at 
second order, the spectral index takes different forms in different frames \cite{kaiser,nozari}. 
This is a consequence of the metrical theory where the FRW metric is conformally transformed from Jordan to Einstein frame.

\begin{figure}[h]
\begin{tikzpicture}
\begin{axis}[
    xmode=log,
    axis lines = box,
    xlabel = The parameter $\xi$,
    ylabel = {The spectral index $n_{s}$},
    ymin=0.950,
    ymax=0.970,
    ytick={0.950,0.960,0.970},
    legend pos=south east,
]
\addplot [
    domain=0.003:10^(3),
    samples=100,
    color=blue,
]
{0.966748 - 1.90723*10^(-8)/(x^(2)) - 0.000211993/x};
\addlegendentry{$\text{AG}$}
\addplot [
    domain=0.003:10^(3),
    samples=100,
    color=red,
    style=dashed,
    ]
    {1 + (74.9653*x^(2))/(-1 + 960*x)^(2) - (32*x)/(-1 +
    960*x)-(16199*x^(2))/((-1 + 960* x)^2 *(1 +960* x)^(2))
    -(3237.89*x^(2))/((-1 + 960*x)^(2)*(1 +960*x))-(192*x)/((-1 + 960*x)*(1 + 960*x))};
\addlegendentry{$\text{GR}$}
\end{axis}
\end{tikzpicture}
\caption{The spectral index $n_{s}$ to second order in the GR and the AG for $N=60$ e-foldings. 
Deviation from the first order is significant in GR. In the AG, the slow roll parameters 
are small and corrections are tiny.  }
\label{fig: second order spectral index}
\end{figure}
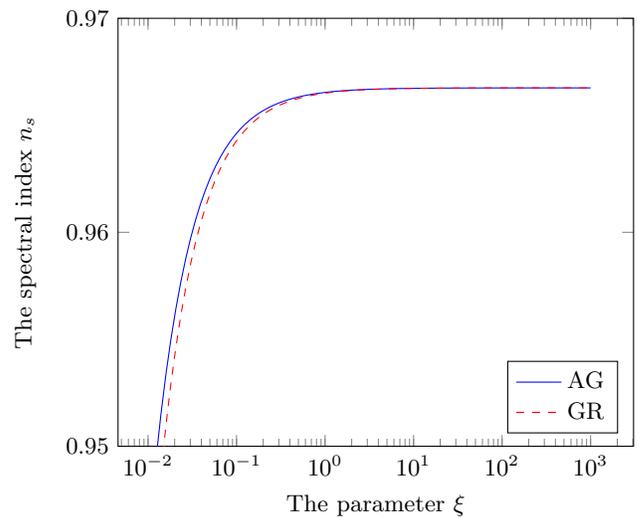

The form of the spectral index (\ref{second order spectral index}) shows that deviations from 
first order is tiny for small slow roll parameters, this is the case of affine inflation where 
this parameters are decaying exponentially. However significant deviation from first order appears in 
GR as it is illustrated in Figure~\ref{fig: second order spectral index}. Unlike the first order case, 
AG does not show much differences from GR.

Last but not least, the tensor-to-scalar ratio is given by
$r \equiv \Delta^{2}_{t}/\Delta^{2}_{s}=16\epsilon $, 
where $\Delta^{2}_{t}$ and $\Delta^{2}_{s}$ are the power spectra of the tensor and scalar fluctuations respectively, 
created by inflation. In AfI, this quantity takes the form
\begin{eqnarray}
r\simeq \frac{2}{\xi N^{2}}.
\end{eqnarray}
It is clear that this ratio is very small. 
For the range given above; $\xi \gtrsim 3.12 \times 10^{-2} $ for $60$ e-foldings, 
this ratio has an upper bound
\begin{eqnarray}
r \lesssim 1.7 \times 10^{-2},
\end{eqnarray}
showing a small amount of tensor perturbations which is in the range of the observed value \cite{planck}. However, a large $\xi$ produces a very tiny ratio.

Planck data constraint on the power spectrum of the primordial perturbations 
generated during inflation is given by \cite{planck}
\begin{eqnarray}
\frac{H^{2}}{8\pi^{2} \epsilon M^{2}_{Pl}} \simeq 2.4 \times 10^{-9},
\end{eqnarray}
which leads to
\begin{eqnarray}
\frac{\lambda}{\xi}   \simeq 7.8 \times 10^{-11}.
\end{eqnarray}

For small parameter $\xi$, this small ratio requires a very small $\lambda$ leading to an extreme fine tuning. 
However, a natural value of $\lambda$ can be obtained here for a significantly
large $\xi$. This case is permitted in our model where the spectral index takes the value
$n_{s} \simeq 0.97$. The results which arise from large non-minimal coupling parameter $\xi$ are 
equivalent to the those obtained in Ref~\cite{bauer}. We believe that future observational 
constraints on the parameter $\xi$ will lead to precise differences between standard inflation based on GR and AfI based on AG.

We conclude this section by summarizing our results in Table~\ref{tab:table3}. It describes the inflationary regime driven 
by non-minimally coupled inflaton in the frameworks of affine gravity and general relativity.

\begin{table}[h]
\begin{ruledtabular}
\begin{tabular}{cccc}
\textrm{}&
\textrm{Einstein frame(GR)}&
\textrm{AG}\\
\colrule
\, & \, & \, & \, \\
Parameter \\ $\xi$ & $\xi \gtrsim 6.25\times 10^{-3}$ & $\xi \gtrsim 3.12 \times 10^{-2}$ \\ 
\, & \, & \, \\
$\phi\left(\varphi\right)$ & $\frac{M_{Pl}}{\sqrt{\xi}} \exp \left(\sqrt{\frac{\xi}{1+6\xi}}\frac{\varphi}{M_{Pl}} \right)$ & $\frac{M_{Pl}}{\sqrt{\xi}}\sinh\left(\frac{\sqrt{\xi}}{M_{Pl}}\varphi\right)$\\
\, & \, & \, \\
$\varphi_{i}/M_{Pl}$ & $\sqrt{\frac{1+6\xi}{\xi}}\ln\left( \sqrt{\frac{8\xi N}{1+6\xi}}\right)$ & $\ln\left(32\xi N \right)/2\sqrt{\xi}$  \\
\, & \, & \, \\
$\varphi_{f}/M_{Pl}$ &$\sqrt{\frac{1+6\xi}{16 \xi}}\ln\left( \frac{8\xi}{1+6\xi}\right)$ & $\ln\left(128\xi \right)/4\sqrt{\xi}$ \\
\end{tabular}
\end{ruledtabular}
\caption{\label{tab:table3}%
Inflationary regime predictions from AG and GR (Einstein frame) that correspond to the first order spectral index $ 0.960 \lesssim n_{s} \lesssim 0.970$. The function $\varphi\left(\phi\right)$ is given here for $\xi >1$, in this case and for larger $\xi$, the fields are below the Planck mass in AG. }
\end{table}

\section{\label{sec:level5} Conclusion}

{Affine gravity, since its first formulation by Eddington, Schr\"{o}dinger and Einstein \cite{eddington}, has remained for decades as a mathematical formulation that lacks concrete physical and cosmological interpretations. The present work may therefore be considered as a modest attempt to utilize the affine gravity for inflationary phase of the Universe. It turns out that the theory provides a viable setup for inflation even if it is equivalent to GR in certain cases (the minimal coupling). This feature stems from the structure of the invariant actions which requires scalar fields to take {non-zero} values. This feature, which is necessary to drive inflation, is a useful aspect of AG. Another important feature of AG is that, it provides a geometric frame in which the generated metric tensor is unique (no Einstein or Jordan frames). This makes the minimal and non-minimal coupling theories in the AG as equivalent descriptions

In this work, we first studied minimally and non-minimally coupled scalar fields
comparatively in the GR and in the AG. We have revealed a number of interesting features
in both cases. The scalar field is required to have non-vanishing potential energy
density in the AG. In effect, energy-momentum tensor of the vacuum is found to
define a metric tensor {\it a posteriori} as an integration constant of 
the equations of motion.  

Another point concerns transition from minimal to non-minimal coupling.
It turns out that, unlike the minimal case, the non-minimal coupling in AG differs from the GR. 
The differences stem from both, the improved energy-momentum tensor and the modified equation of the field $\phi$. 
We have shown that the improved energy momentum tensor depends on the potential of the scalar field rather 
than derivatives of the field $\phi$ as in GR. This is a consequence of the linearity of the 
Ricci tensor in first derivative of the affine connection. 

We have also shown that the  transformation from non-minimal to minimal coupling 
is simply obtained through the scalar field re-definition. This shows that there is 
only one frame in which affine gravity is formulated. This is arguably clear since there 
is only one generated metric tensor. This means that Jordan and Einstein frames of GR 
have no correspondent in AG.

In the final stage of the paper, we have presented a detailed study of 
the primordial inflaton in a unique FRW spacetime metric and we have 
shown that an inflationary regime arise naturally for slowly moving-fields. 
We have discussed the possible values of the non-minimal coupling parameter 
$\xi$ based on the measured spectral index. The study also showed that unlike the 
standard inflation based on GR, AfI produces a small tensor-to-scalar ratio.
Future observations may reach the sensitivity to distinguish between these
models.

\section*{acknowledgments}
This work is supported in part by the T{\"U}B{\.I}TAK grant 115F212. H. A thanks to Enrico Pajer for discussions on affine gravity. We are thankful to the anonymous referee for the critical and constructive comments.

\bibliography{apssamp}

\end{document}